\documentstyle[12pt]{article}

\newcounter{eq}
\newcounter{sc}

\def\overleftrightarrow#1{\vbox{\ialign{##\crcr
 $\leftrightarrow$\crcr\noalign{\kern-1pt\nointerlineskip}
 $\hfil\displaystyle{#1}\hfil$\crcr}}}

\newlength{\minitwocolumn}
\begin{document}

\begin{flushright}
DPUR/TH/30\\
April, 2012\\
\end{flushright}
\vspace{20pt}

\pagestyle{empty}
\baselineskip15pt

\begin{center}
{\large\bf Velocity of Light in Dark Matter with Charge
\vskip 1mm }

\vspace{20mm}
Ichiro Oda \footnote{E-mail address:\ ioda@phys.u-ryukyu.ac.jp
}

\vspace{5mm}
           Department of Physics, Faculty of Science, University of the 
           Ryukyus,\\
           Nishihara, Okinawa 903-0213, Japan.\\

\end{center}


\vspace{5mm}
\begin{abstract}
We propose an interesting mechanism to reconcile the recent experiments
of the Michelson-Morley type and slowdown of the velocity of light in dark matter
with a fractional electric charge when the index of refraction of dark matter depends 
on the frequency of a photon. After deriving the formula for the velocity of light 
in a medium with the index of refraction $n(\omega)$ in a relativistic regime, 
it is shown that the local anisotropy of the light speed is proportional to the second order in
$n(\omega) - 1$. This result implies that the experiments of the Michelson-Morley type
do not give rise to a stringent constraint on the slowdown of the velocity of light in 
dark matter with electric charge. 
\end{abstract}

\newpage
\pagestyle{plain}
\pagenumbering{arabic}


\rm
\section{Introduction}

It is nowadays well known that a large portion, about $23$ percentages, of the total
energy density in the present universe is occupied by unknown material,
what we call "dark matter". The particles, of which dark matter is made, are usually thought
to be weakly interacting massive ones, in other words, WIMPs, which are absent in the Standard 
Model (SM). These hypotheoretical particles are supposed to have mass of order ranging 
from GeV to TeV, and cross section of their annihilation process is comparable to that 
of the weak interaction. Thus they do not completely annihilate in the course of evolution 
of the universe and their mass density in the present universe could be of the order of 
the critical density $\rho_c \approx 0.52 \times 10^{-5} GeV/cm^3$ (more precisely, 
$\rho_{DM} \approx 1.2 \times 10^{-6} GeV/cm^3$ averaged over large distance scales) \cite{Rubakov}. 

The difficulty of terrestrial experiments for detecting dark matter by direct 
or indirect methods makes it impossible to clarify its physical properties and therefore
answer various questions associated with dark matter such as the mechanism of the generation 
in the early universe. On the other hand, the density of dark matter in clusters of galaxies is 
determined by various methods of measurement of the gravitational potential.  For instance,
mass distribution in a cluster is obtained by the method of gravitational lensing. 
The result is that most of the mass is due to dark matter distributed smoothly over
the cluster. Put differently, dark matter, like the usual matter, is more dense in galaxies and
it appears as if luminous matter is embedded into the cloud of dark matter
of larger size-galactic halo. It is easy to understand this concentration of mass of dark matter 
near clusters of galaxies from a physical viewpoint since the gravitational interaction attracts 
dark matter near the clusters compared to empty outer space.  

Given our current ignorance of the dark sector in the universe, it seems prudent not to restrict
our interests only in WIMPs, though they must be most plausible, and open our mind to many possibilities.
Indeed, though the particles consisting of WIMPs carry no electric charge, the notion of charged massive
particles (CHAMPs) has also appeared periodically \cite{Goldberg}-\cite{Davidson2}. Since CHAMPs, 
which normally carry a full unit of electric charge, receive stringent constraints from searches 
for heavy hydrogen to direct detection in the underground experiments, the possibility of CHAMPs 
carrying a fractional charge has been recently investigated \cite{Dubovsky, Gardner}.   

In this article, we shall assume that the earth itself is embedded into a medium of
dark matter with a fractional electric charge. Then, it is quite natural to think that 
the existence of such a dark matter manifests itself the index of refraction $n(\omega)$ depending
on the (angular) frequency of light since the dark matter can interact with the gauge field 
in a direct way \cite{Jackson}.\footnote{Precisely speaking, an electrically-neutral particle propagating 
through matter is also refracted if it couples to a charged particle even indirectly. For instance, 
a neutrino can couple to an electron via weak interaction and its refractive property has an important effect
on neutrino propagation when the neutrino is massive \cite{Wolfenstein}.}  
Moreover, we can ask ourselves if the recent experiments of the Michelson-Morley type \cite{Herrmann, Eisele}
shed some light on the possibility of detecting the dark matter on the earth 
through measuring the local anisotropy of velocity of light or not. 

One of motivations behind the present study comes from a recent claim \cite{Frere} that 
in case of the OPERA superluminal neutrinos \cite{OPERA}, such an assumption shows a clear tension 
and might be in conflict with the most recent experiments of the Michelson-Morley type about 
the local anisotropy of velocity of light \cite{Herrmann, Eisele}. In order to relax this tension, 
it is useful to recall that the index of refraction of a medium is in general dependent 
on the frequency of light. If we take account of this fact together with its property under 
the Lorentz transformation in a proper manner, a term depending on the frequency 
provides an additional contribution to the velocity of light and then the resulting constraint 
would become consistent with the experimental results of the Michelson-Morley type.

This article is organized as follows: In the next section, we derive a formula of
propagation of light in moving media based on special relativity. In Section 3, it is shown 
that using this formula the local anisotropy of the light speed is proportional to the second order in
$n(\omega) - 1$. Section 4 is devoted to discussion.

\section{Velocity of propagation of light in moving media}

In this section, we review special relativity \cite{Einstein}, 
in particular, on the basis of both the addition law of relativistic velocities 
and the invariance of phase of a plane wave, we shall derive the velocity formula for 
propagation of light in moving media.

It is known that the Lorentz transformation for inertial coordinates can be
used to derive the addition law of relativistic velocities. To show this fact
explicitly, let us consider a physical setup where we have a moving point $P$
whose three-dimensional velocity vector $\vec{u}'$ has spherical coordinates 
$(u', \theta', \phi')$ in the inertial frame $K'$, and the frame $K'$ is
moving with velocity $\vec{v} = c_l \vec{\beta}$ in the
direction of the $x_1$ axis with respect to the inertial reference frame $K$. 
Here $c_l$ is a universal limiting speed, in other words, the maximum speed of 
all physical entities \cite{Ignatowsky}-\cite{Oda3}.

Then, one wishes to calculate the velocity $\vec{u}$ of the
point $P$ as seen from the inertial frame $K$. To do so, let us first note that
the Lorentz transformation takes the form for the differential expressions
\begin{eqnarray}
dx_0 &=&  \gamma (dx_0' + \beta dx_1'), \nonumber\\
dx_1 &=&  \gamma (dx_1' + \beta dx_0'), \nonumber\\
dx_2 &=&  dx_2', \hspace{2.0em} dx_3 = dx_3', 
\label{Velo-transf}
\end{eqnarray}
where we have defined $\gamma \equiv \frac{1}{\sqrt{1 - (\frac{v}{c_l})^2}}$. 
Since the components of velocity are defined by $u_i' = c_l \frac{dx_i'}{dx_0'}$ 
and $u_i = c_l \frac{dx_i}{dx_0}$, they transform as
\begin{eqnarray}
u_1 =  \frac{u_1' + v}{1 + \frac{v u_1'}{c_l^2}}, 
\hspace{2.0em}
u_2 =  \frac{u_2'}{\gamma(1 + \frac{v u_1'}{c_l^2})}, 
\hspace{2.0em}
u_3 =  \frac{u_3'}{\gamma(1 + \frac{v u_1'}{c_l^2})}. 
\label{Lo-transf0}
\end{eqnarray}
For later convenience, let us rewrite this transformation to a more general form
\begin{eqnarray}
u_{||} =  \frac{u_{||}' + v}{1 + \frac{\vec{v} \cdot \vec{u}'}{c_l^2}}, 
\hspace{2.0em}
\vec{u}_\bot = \frac{\vec{u}_\bot'}{\gamma(1 + \frac{\vec{v} 
\cdot \vec{u}'}{c_l^2})},
\label{Lo-transf}
\end{eqnarray}
where $u_{||}$ and $\vec{u}_\bot$ indicate the components of velocity 
parallel and perpendicular to $\vec{v}$, respectively. Moreover, it turns out
that the magnitude $u$ of $\vec{u}$ and its polar angles in the two inertial
frames are related as follows:
\begin{eqnarray}
\phi &=&  \phi',  \nonumber\\
\tan \theta &=& \frac{u' \sin \theta'}{\gamma(u' \cos \theta' + v)}, \nonumber\\ 
u &=& \frac{\sqrt{u'^2 + v^2 + 2 u'v \cos \theta' - (\frac{u'v \sin \theta'}{c_l})^2 }}
{1 + \frac{u'v}{c_l^2} \cos \theta'}.
\label{u-comp}
\end{eqnarray}
The inverse transformation for $\vec{u}'$ in terms of $\vec{u}$
can be easily obtained by interchanging primed and unprimed quantities and 
simultaneously changing the sign of $v$.

In what follows, for simplicity, let us focus on the case of the parallel velocities,
$\theta' = 0$. In this case, the magnitude of $u$ takes the form
\begin{eqnarray}
u = \frac{u' + v}{1 + \frac{u'v}{c_l^2}}.
\label{u}
\end{eqnarray}
If one particularly sets $u' = c_l$, this expression reads $u = c_l$, which simply means 
the postulate of a universal limiting speed, which is an alternative fundamental principle 
for the principle of invariant speed of light in special relativity \cite{Ignatowsky, Frank,
Mermin, Terletskii}.
  
Next, let us choose $u_{||}' = \frac{c_l}{n(\omega')}$, thereby giving rise to
\begin{eqnarray}
u_{||} =  \frac{\frac{c_l}{n(\omega')} + v}{1 + \frac{v}{n(\omega') c_l}}, 
\label{u||}
\end{eqnarray}
where $\omega'$ is the (angular) frequency of light in the inertial frame $K'$ and
$n(\omega')$ is the index of refraction of a medium. The important point is
that the index of refraction of a medium in general depends on the magnitude
of the frequency of the photon.

In order to determine $n(\omega')$, we make use of the fact that the phase of
a plane wave is an invariant quantity under the Lorentz transformation since
the phase $\varphi$ can be identified with the mere counting of wave crests in a 
wave train, an operation that must be the same in every inertial frame
\begin{eqnarray}
\varphi \equiv k_\mu x^\mu = \omega t - \vec{k} \cdot \vec{x} 
= \omega' t' - \vec{k}' \cdot \vec{x}'. 
\label{phase}
\end{eqnarray}
With the frequencies $\omega = c_l k_0$ and $\omega' = c_l k_0'$, the Lorentz transformation
for the wave-number vector $k_\mu$ reads
\begin{eqnarray}
k_0' &=&  \gamma (k_0 - \vec{\beta} \cdot \vec{k}), \nonumber\\
k_{||}' &=&  \gamma (k_{||} - \beta k_0), \nonumber\\
\vec{k}_\bot' &=&  \vec{k}_\bot. 
\label{k-transf}
\end{eqnarray}
For light waves, $|\vec{k}| = k_0$ and $|\vec{k}'| = k_0'$, so using $\theta = 0$ coming from
$\theta' = 0$ via Eq. (\ref{u-comp}) we have the relation
\begin{eqnarray}
\omega' =  \gamma (1 - \beta) \omega = \sqrt{\frac{1-\beta}{1+\beta}} \omega
\approx (1 - \beta) \omega, 
\label{omega-transf}
\end{eqnarray}
where the last relation holds when $\beta \equiv \frac{v}{c_l}$ is small.
Thus, the index of refraction of a medium transforms to first order in $\beta$
as 
\begin{eqnarray}
n(\omega') =  n(\omega) - \beta \omega \frac{dn}{d\omega}. 
\label{n}
\end{eqnarray}
Plugging this expression into Eq. (\ref{u||}) (and taking $\pm v$ into consideration at the
same time), it is easy to show that for medium flow at a speed $v$ parallel or antiparallel to the path 
of the light, the speed of the light, as observed in the laboratory ($K$-frame), is given to first order 
in $v$ by
\begin{eqnarray}
u_{||}^\pm \approx  \frac{c_l}{n(\omega)} \pm v \left[ 1 - \frac{1}{n(\omega)^2} 
+ \frac{\omega}{n(\omega)^2} \frac{dn(\omega)}{d \omega} \right], 
\label{u||+-}
\end{eqnarray}
where superscripts $+$ and $-$ on $u_{||}$ denote the speed parallel and antiparallel to the light path,
respectively. Formula (\ref{u||+-}) is the main result of this section and will be
used in the next section.

\section{The recent experiments of Michelson-Morley type}

We wish to apply the formula (\ref{u||+-}) to the problem that the recent experiments of 
the Michelson-Morley type could or could not give us useful information on detection of dark matter
with electric charge. 

Before doing that, let us present some ideas behind our theory. Search for dark matter 
is underway, but no conclusive evidence has been obtained thus far. A direct way for the search
is carried out in experiments trying to detect energy deposition in a detector caused by
elastic scattering of dark matter off a nucleus. Besides the direct search, there are 
experiments to search for dark matter particles indirectly, which include the search
for products of dark matter annihilation. One of the most promising indirect ways is to
search for monoenergetic photons which are emitted in two-body annihilation processes
$X + \bar X \rightarrow \gamma + \gamma, X + \bar X \rightarrow Z^0 + \gamma$ where $X, \bar X$
describe a dark matter particle and its antiparticle, respectively. By the crossing
symmetry, these processes read $X + \gamma \rightarrow X + \gamma, X + \gamma \rightarrow X + Z^0$.
If the photon energy is small compared to the weak energy scale (i.e., the energy of the photon
beam is below the threshold for production of a single $Z^0$), only the former process is allowed to
occur and it is nothing but the elastic Compton scattering. 

At this stage, let us recall that there is a well-known connection between the index of refraction 
and the Compton scattering amplitude where the standard formula takes the form \cite{Gell-Mann} 
\begin{eqnarray}
n(\omega) = 1 +  \frac{2 \pi c_l^2 N F(\omega)}{\omega^2}, 
\label{index}
\end{eqnarray}
where $N$ is the number of scattering centers per unit volume and $F(\omega)$ is the forward Compton
scattering amplitude for the photon scattering off the medium which is a function of the (angular) frequency
$\omega$ of a photon. Put differently, in this approach the dark matter with electric charge plays a role
of a medium for the photons and provides a dispersive effect on light propagation.

Since we assume in this article that dark matter carries a fractional electric charge associated with 
$U(1)_{EM}$ symmetry, it must be a stable Dirac particle with spin $\frac{1}{2}$ like protons.
Then, the amplitude for the forward scattering of a photon from a polarization state $\vec{e}_1$ 
to the one $\vec{e}_2$ by the dark matter must be of the general form \cite{Gell-Mann}
\begin{eqnarray}
F(\omega) = \psi_f^* \left[ F_1(\omega) \vec{e}_2^* \cdot \vec{e}_1 
+  i F_2(\omega) \vec{\sigma} \cdot (\vec{e}_2^* \times \vec{e}_1) \right] \psi_i, 
\label{F}
\end{eqnarray}
where $\psi_i (\psi_f)$ is the wave function of the initial (final) dark matter 
and $\vec{\sigma}$ is the spin matrix of the dark matter. In general, both $F_1(\omega)$ and $F_2(\omega)$ 
are complex and have dispersive and absorptive parts. If one averages over dark matter spins in the
amplitude one is left only with $F_1(\omega)$. The amplitudes $F_1(\omega)$ and $F_2(\omega)$ are
separable if one can do experiments with polarized dark matters: $F_1(\omega)$ corresponds to
parallel and $F_2(\omega)$ to perpendicular linear polarization vectors of the initial and final photons,
respectively. Since we are interested in the spin-averaged forward amplitude, we will focus on only
$F_1(\omega)$ henceforth. (We therefore define $F = F_1$.) 

With the assumption of causality and analyticity for the forward scattering amplitude $F(\omega)$,
the once-subtracted dispersion relation is given by
\begin{eqnarray}
Re F(\omega) = Re F(0) + \frac{2 \omega^2}{\pi} P \int_{\omega_0}^\infty d \omega'
\frac{Im F(\omega')}{\omega' ( \omega'^2 - \omega^2)}, 
\label{Dispersion}
\end{eqnarray}
where $P$ denotes the principal value and $\omega_0$ is the threshold for producing a single 
dark matter which is approximately taken to be $\omega_0 \approx M_X c_l^2 \approx 100 GeV$. Using the optical theorem, 
the imaginary part of the forward elastic scattering amplitude is related to the total cross section 
$\sigma(\omega)$ by
\begin{eqnarray}
\sigma (\omega) = \frac{4 \pi c_l}{\omega} Im F(\omega). 
\label{Optical}
\end{eqnarray}
Furthermore, it is known that the forward elastic scattering amplitude at $\omega =0$ is real and given by
the Thomson formula
\begin{eqnarray}
Re F(0) = - \frac{(\varepsilon e)^2}{M_X c_l^2}, 
\label{Thomson}
\end{eqnarray}
where $\varepsilon e$ and $M_X$ are respectively the charge and mass of the dark matter particle. 
Thus, one can rewrite $Re F(\omega)$ in Eq. (\ref{Dispersion}) to the form
\begin{eqnarray}
Re F(\omega) =  - \frac{(\varepsilon e)^2}{M_X c_l^2} + \frac{\omega^2}{2 \pi^2 c_l} 
P \int_{\omega_0}^\infty d \omega' \frac{\sigma(\omega')}{\omega'^2 - \omega^2}. 
\label{Dispersion2}
\end{eqnarray}
Substituting this expression into the standard formula (\ref{index}), we obtain
\begin{eqnarray}
Re n(\omega) - 1 =  - \frac{2 \pi N (\varepsilon e)^2}{M_X} \frac{1}{\omega^2}
+ \frac{c_l N}{\pi} P \int_{\omega_0}^\infty d \omega' \frac{\sigma(\omega')}{\omega'^2 - \omega^2}. 
\label{Re-n}
\end{eqnarray}
Since $Im n(\omega)$ does not appear any more and $n(\omega)$ in the previous section is in fact 
equivalent to $Re n(\omega)$ in this section, for simplicity we rewrite $Re n(\omega)$ as $n(\omega)$
from now on.
   
Now let us consider the case of the low photon energy of $\omega \ll \omega_0$.\footnote{In the most 
recent Michelson-Morley experiments \cite{Herrmann, Eisele}, photons with the order $100 THz$ 
are utilized, which corresponds to the order $1 eV$.}  It is then sufficient to limit ourselves to 
an expansion up to the constant order in $\omega^2$ \footnote{Expanding in powers of $\omega$ is known
to be a useful tool for obtaining various interesting low-energy cross sections in the analysis of
the Compton scattering \cite{Griesshammer}.}
\begin{eqnarray}
\delta(\omega) \equiv n(\omega) - 1 \approx \frac{\delta_0}{\omega^2}  + \delta_2, 
\label{Delta}
\end{eqnarray}
where $\delta_0, \delta_2$ are defined as 
\begin{eqnarray}
\delta_0 =  - \frac{2 \pi N (\varepsilon e)^2}{M_X},   \hspace{2.0em}
\delta_2 = \frac{c_l N}{\pi} P \int_{\omega_0}^\infty d \omega' \frac{\sigma(\omega')}{\omega'^2}. 
\label{delta0-2}
\end{eqnarray}
Note that observations of $\gamma$-ray burst (GRB) emission from radio energies to tens of GeV \cite{Abdo} 
require us that the energy dependence of the speed of the photon is so small that $\delta_0$ must be very tiny. 
Moreover, the recent ICARUS result \cite{ICARUS} implies that $\delta_2$ are very small as well.\footnote{Although 
$E = \hbar \omega$ corresponds to the energy of a photon, we will regard it as a typical energy scale controlling 
the system under consideration, in other words, the energy of the ICARUS neutrinos.} 

However, this smallness is not a real problem for the present purpose. The key point is that 
if we regard dark matter as a kind of usual medium for light, the frequency dependence must be always 
given by (\ref{Delta}) at least for the low energy photons.
Then, an important relation, which will be utilized shortly, can be obtained
\begin{eqnarray}
\omega \frac{d n(\omega)}{d \omega} = -2 (n - 1 - \delta_2). 
\label{index2}
\end{eqnarray}

Next, following the procedure in Ref. \cite{Frere}, let us calculate the local anisotropy of
the speed of light whose experimental upper bound has been already obtained in the experiments of 
the Michelson-Morley type \cite{Herrmann, Eisele}
\begin{eqnarray}
\left(\frac{\Delta c}{c}\right)_{Exp} \approx 1 \times 10^{-17}. 
\label{Exp-anisotropy}
\end{eqnarray}
In these experiments, the parallel average velocity is measured
\begin{eqnarray}
c_{||} = \frac{2L}{\frac{L}{u_{||}^+} + \frac{L}{u_{||}^-}}. 
\label{Av-velocity}
\end{eqnarray}
Using Eqs. (\ref{u||+-}) and (\ref{index2}), this quantity can be evaluated to be
\begin{eqnarray}
c_{||} = \frac{c_l}{n} - \frac{v^2}{c_l} \frac{\left[(n-1)^2 + 2 \delta_2 \right]^2}{n^3}. 
\label{Av-velocity2}
\end{eqnarray}
Then, the theoretical local anisotropy of the speed of light is calculated to be
\begin{eqnarray}
\frac{\Delta c}{c} = \frac{\frac{c_l}{n(\omega)} - c_{||}}{\frac{c_l}{n(\omega)}}
\approx 4 \left(\frac{v}{c_l}\right)^2 \delta_2^2
\approx 4 \left(\frac{v}{c_l}\right)^2 \delta^2, 
\label{The-anisotropy}
\end{eqnarray}
where we have used $\delta \approx 0, \beta \ll 1$ and $\delta \approx \delta_2$. 
It is worthwhile to notice that compared to the result in Ref. \cite{Frere} which is proportional 
to the linear order in $\delta$, our result (\ref{The-anisotropy}) becomes proportional to 
the second order in $\delta$ since we have taken account of the frequency dependence and 
the Lorentz transformation of the index of refraction $n(\omega)$.

In order to show concretely that Eq. (\ref{The-anisotropy}) yields a very weak constraint
on the local anisotropy of the speed of light, let us apply it to the OPERA result \cite{OPERA}  
together with the SN1987A result \cite{Hirata}. Here note that the OPERA result has been recently 
refuted by the ICARUS group \cite{ICARUS}. The reason why we use the OPERA result is two-fold.
First, Ref. \cite{Frere} has discussed the OPERA result and shown that there is some tension 
between the recent Michelson-Morley experiments and the models based on CHAMPs owing to a large 
value of $\delta$ in the OPERA result. But we wish to show that even in this large value of $\delta$ 
there appears no tension between them since (\ref{The-anisotropy}) becomes proportional to the second order in $\delta$
in our calculation. And the second reason is that even if the OPERA main result, in particular 
the size of $\delta$ might be wrong, the other result, i.e., $\delta$ is independent of the neutrino energy, 
is consistent with that of GRB emission \cite{Abdo} and seems to be true.  

The result of SN1987A implies that the velocity of a neutrino is almost the same as a universal 
limiting speed
\begin{eqnarray}
v_\nu \approx c_l. 
\label{SN1987A}
\end{eqnarray}
Then, we can obtain the following relation
\begin{eqnarray}
\frac{v_\nu - \frac{c_l}{n}}{\frac{c_l}{n}} \approx \frac{c_l - \frac{c_l}{n}}{\frac{c_l}{n}}
= n - 1 = \delta \approx 2.37 \times 10^{-5}, 
\label{OPERA}
\end{eqnarray}
where we have used the OPERA results in the last step. The OPERA result suggests that $\delta$ is
independent of the neutrino energy, so this equation implies 
\begin{eqnarray}
\delta_2 \approx \delta \approx 2.37 \times 10^{-5}. 
\label{delta2}
\end{eqnarray}
Substituting this value and the equatorial rotation speed of the earth at the observed place, $\beta \equiv \frac{v}{c_l}
= 1.55 \times 10^{-6}$ \cite{Frere} into Eq. (\ref{The-anisotropy}), one arrives at the result
\begin{eqnarray}
\left(\frac{\Delta c}{c}\right)_{OPERA} \approx 4.6 \times 10^{-21}. 
\label{The-anisotropy2}
\end{eqnarray}
Note that this value is much smaller than the experimental upper bound (\ref{Exp-anisotropy}),
so it is not in conflict with the recent experimental results of the Michelson-Morley type. 
Of course, our result does not refute the result obtained in Ref. \cite{Frere} directly since
in our approach at hand the index of refraction depends on the frequency of a photon whereas
in the approach considered in Ref. \cite{Frere} the index of refraction is assumed to be independent 
of the photon frequency. Actually, when dark matter does not carry an electric charge like a
neutrino, Eq. (\ref{delta0-2}) means $\delta_0 = 0$, thereby making it impossible to get 
Eq. (\ref{index2}) owing to $\omega \frac{d n(\omega)}{d \omega} = 0$.

For completeness, let us move on to a general case of the arbitrary photon energy.
In this case, we can proceed the argument in a perfectly similar way to the case of the
low energy photons. The relation (\ref{index2}) is now replaced with
\begin{eqnarray}
\omega \frac{d n(\omega)}{d \omega} = -2 (n - 1 - \tilde \delta), 
\label{index3}
\end{eqnarray}
where $\tilde \delta$ is defined as
\begin{eqnarray}
\tilde \delta =  \frac{c_l N}{\pi} P \int_{\omega_0}^\infty d \omega' 
\frac{\omega'^2 \sigma(\omega')}{(\omega'^2 - \omega^2)^2}. 
\label{tilde delta}
\end{eqnarray}
Then, it turns out that local anisotropy of the speed of light is exactly evaluated to be
\begin{eqnarray}
\frac{\Delta c}{c} = \left(\frac{v}{c_l}\right)^2 
\frac{[ (n-1)^2 + 2 \tilde \delta ]^2}{n^2}. 
\label{The-anisotropy3}
\end{eqnarray}
With the reasonable assumptions as before, $\delta \equiv n - 1 \approx 0, \beta \ll 1$ and 
$\delta \approx \tilde \delta$, we can obtain a similar result to Eq. (\ref{The-anisotropy})
\begin{eqnarray}
\frac{\Delta c}{c} \approx  4 \left(\frac{v}{c_l}\right)^2 \tilde \delta^2
\approx  4 \left(\frac{v}{c_l}\right)^2 \delta^2. 
\label{The-anisotropy4}
\end{eqnarray}
Compared to the low energy case, a slight modification here appears in the $\tilde \delta$ 
which is dependent on the photon frequency as can be seen in (\ref{tilde delta}). However, 
except this fact, the essential feature remains unchanged so the conclusion obtained in the case of 
the low enery photons still holds even in this general case .

\section{Discussion}

Since the introduction of dispersion relations into elemetary particle physics,
originally within the context of quantum field theory \cite{Gell-Mann}, a large number of 
literatures have grown up on their theoretical basis, on extensions and applications to new
processes, and on their comparison with experiment with great success. The most
advantageous point is that dispersion relations are very universal in the sense that
they are formulated only on the basis of fundamental principles of quantum field theory
such as causality, the Lorentz invariace and analyticity. 

As a simple application of the dispersion relations, in this article, we have studied 
the velocity of light in dark matter with a fractional electric charge and found that 
the recent experiments of the Michelson-Morley type \cite{Herrmann, Eisele} do not provide 
a stringent condition on the slowdown of the velocity of light if the index of refraction 
of dark matter depends on the frequency of a photon. The experiments of the Michelson-Morley 
type have played a critical role in proving the non-existence of the $\it{aether}$, 
i.e., a medium that was once supposed to fill all space and to support the propagation 
of electromagnetic waves, and might reveal a breakdown of the Lorentz invariance at the 
Planck scale in future. But they do not seem to shed some light on the change of the velocity 
of light in dark matter with charge. Perhaps, if the ICARUS result is correct, 
this situation will not change even in the future experiments of the Michelson-Morley 
type.\footnote{It is predicted in Ref. \cite{Herrmann} that $\left(\frac{\Delta c}{c}\right)_{Exp} 
\approx 1 \times 10^{-20}$ regime is possible.}  

Nevertheless, the experiments using the photons will provide important insight into 
the properties of CHAMPs in future. For instance, polarized Compton scattering will yield
information on spin-structure of the dark matter. Moreover, if one increases the
energy of the incoming photon beam, one can probe the internal structure of the
dark matter \cite{Griesshammer}.  

Finally, we wish to close this article by mentioning two future problems to be solved. 
An interesting application of the refractive effect in our model is calculation of the
force acting on a macroscopic body when dark matter flux passes through. Simple calculation
turns out to give a force proportional to $|n(\omega) -1|$ which could be measured in the 
Eotvos-Dicke experiment. Furthermore, the refractive effect in a high-temperature
background such as in the early universe deserves study in future. It is of interest to notice 
that the presence of the electric charge might ensure the stability of the dark matter 
from another angle.

\begin{flushleft}
{\bf Acknowledgements}
\end{flushleft}

This work is supported in part by the Grant-in-Aid for Scientific 
Research (C) No. 22540287 from the Japan Ministry of Education, Culture, 
Sports, Science and Technology.


\end{document}